\documentclass[preprint,aps]{revtex4}

\usepackage{graphicx}
\usepackage{dcolumn}
\usepackage{bm}

\begin{document}

\providecommand{\ignore}[1]{}
\newcommand{\comment}[1]{}
\newcommand{\itcomment}[1]{}
\renewcommand{\comment}[1]{{\color[rgb]{0,1,0}[#1]}}
\renewcommand{\itcomment}[1]{\emph{#1}}
\newcommand{\ket}[1]{{{|}{#1}\rangle}}
\newcommand{\bra}[1]{{\langle{#1}{|}}}
\newcommand{\braket}[2]{{\langle{#1}{|}{#2}\rangle}}
\title{Transport quantum logic gates for trapped ions}

\author
{D. Leibfried, E. Knill, C. Ospelkaus and D. J. Wineland}
\affiliation{National Institute of
 Standards and Technology, 325 Broadway, Boulder, CO 80305, USA}
\date{\today}

\begin{abstract}
Many efforts are currently underway to build a device capable of
large scale quantum information processing (QIP). Whereas QIP has
been demonstrated for a few qubits in several systems, many
technical difficulties must be overcome in order to construct a
large-scale device. In one proposal for large-scale QIP, trapped
ions are manipulated by precisely controlled light pulses and moved
through and stored in multizone trap arrays. The technical overhead
necessary to precisely control both the ion geometrical
configurations and the laser interactions is demanding. Here we
propose methods that significantly reduce the overhead on laser beam
control for performing single and multiple qubit operations on
trapped ions. We show how a universal set of operations can be
implemented by controlled transport of ions through stationary laser
beams. At the same time, each laser beam can be used to perform many
operations in parallel, potentially reducing the total laser power
necessary to carry out QIP tasks. The overall setup necessary for
implementing transport gates is simpler than for gates executed on
stationary ions. We also suggest a transport-based two-qubit gate
scheme utilizing microfabricated permanent magnets that can be
executed without laser light.
\end{abstract}

\maketitle

\section{Introduction}\label{Sec:Int}

A number of physical implementations have been proposed for quantum
information processing (QIP) \cite{roadmap}. This paper is based on
a proposal where trapped atomic ion qubits are to be held in a large
trap array \cite{wineland98,kielpinski02}. Its implementation
requires transporting ions between separated zones, precise control
of local potentials and, at the same time, precise control of laser
beam pointing, intensity and pulse shape. These requirements create
an imposing overhead of classical control for large trap arrays with
multiple interaction zones. Ion transport is accomplished by
electronically changing the potentials of individual control
electrodes in the trap array \cite{rowe02} and might be realized
with on-board CMOS electronics \cite{kim05}, a technology with a
long and very successful track record for scaling. The situation is
very different for the optics necessary for laser beam control:
Microfabricated beam steering optics and electro-optical devices are
typically still ``one-of-a-kind'' designs with only small numbers
produced and scalability in the context of QIP still to be
demonstrated. This problem is compounded by the wavelengths that are
of interest in QIP with trapped ions, which are typically in the
near UV between 214 nm and 400 nm. In addition, a mature optical
fiber technology does not yet exist for this wavelength range. It is
anticipated that high laser power will be required in QIP with
trapped ions \cite{ozeri06}, so beam-splitters or lossy elements
should be used sparingly. At the same time, fault tolerant
architectures require implementing parallel operations
\cite{divincenzo01}. Miniaturization of the currently used approach
with switched beams, as discussed in \cite{kim05} (for example), is
based on the use of a large number of beamsplitters and control
elements to achieve both parallel and individual control of many
different gate operations. Even if the elements used have little
loss (which is currently hard to achieve in the UV), parallel
operations would magnify the already demanding power requirements.

The purpose of this paper is to show that precise control of the
time-dependent external potentials used to transport ions within a
trap array can replace the requirement for precise temporal control
of laser beam intensity in order to implement universal quantum
computation. In an architecture based on transport, laser beams can
be switched on and off collectively with relaxed requirements on
timing and on/off ratios. Such a scenario may also allow for
efficient use of one and the same laser beam in many parallel
operations, thus achieving parallelism without the need for higher
laser power. Under such circumstances it is even conceivable to
further enhance the available power in laser beams with optical
cavities of modest finesse.

The paper is organized as follows: Section \ref{Sec:BasArc} outlines
the basic architecture and QIP primitives necessary for universal
quantum computation with the proposed scheme. We concentrate on
qubits that are comprised of the hyperfine states of ions, which are
manipulated by stimulated two-photon Raman transitions
\cite{wineland98}, but it is possible to adapt the basic
architecture for ion qubits of a different type. Section
\ref{Sec:RamTra} and \ref{Sec:ParDes} briefly summarize the
necessary Raman laser interactions and the spatial dependence of the
laser beam modes used in subsequent sections. Sections
\ref{Sec:OneQub} and \ref{Sec:TwoQub} outline the details of
one-qubit rotations and two-qubit phase gates implemented by ion
transport through laser beams. Section \ref{Sec:SymRec} discusses
how sympathetic cooling can be incorporated and section
\ref{Sec:Ext} introduces some possible extensions of the scheme,
including a two-qubit gate based on transporting ions over a
periodic array of microfabricated permanent magnets, without the
need for laser beams. Finally we summarize and offer some
conclusions in section \ref{Sec:Con}.

\section{Basic architecture}\label{Sec:BasArc}

The goal of the architecture discussed here is to minimize the
requirements on laser-beam steering, pulse shaping and switching as
much as possible by utilizing temporal control  of potentials
applied to the ions in a multizone trap array. Temporal control is
already needed for efficient transport, separation and recombination
of ions, so, with refinements, we can also employ it for qubit gate
operations.

\begin{figure}[ht]
\centerline{\includegraphics[width=10cm]{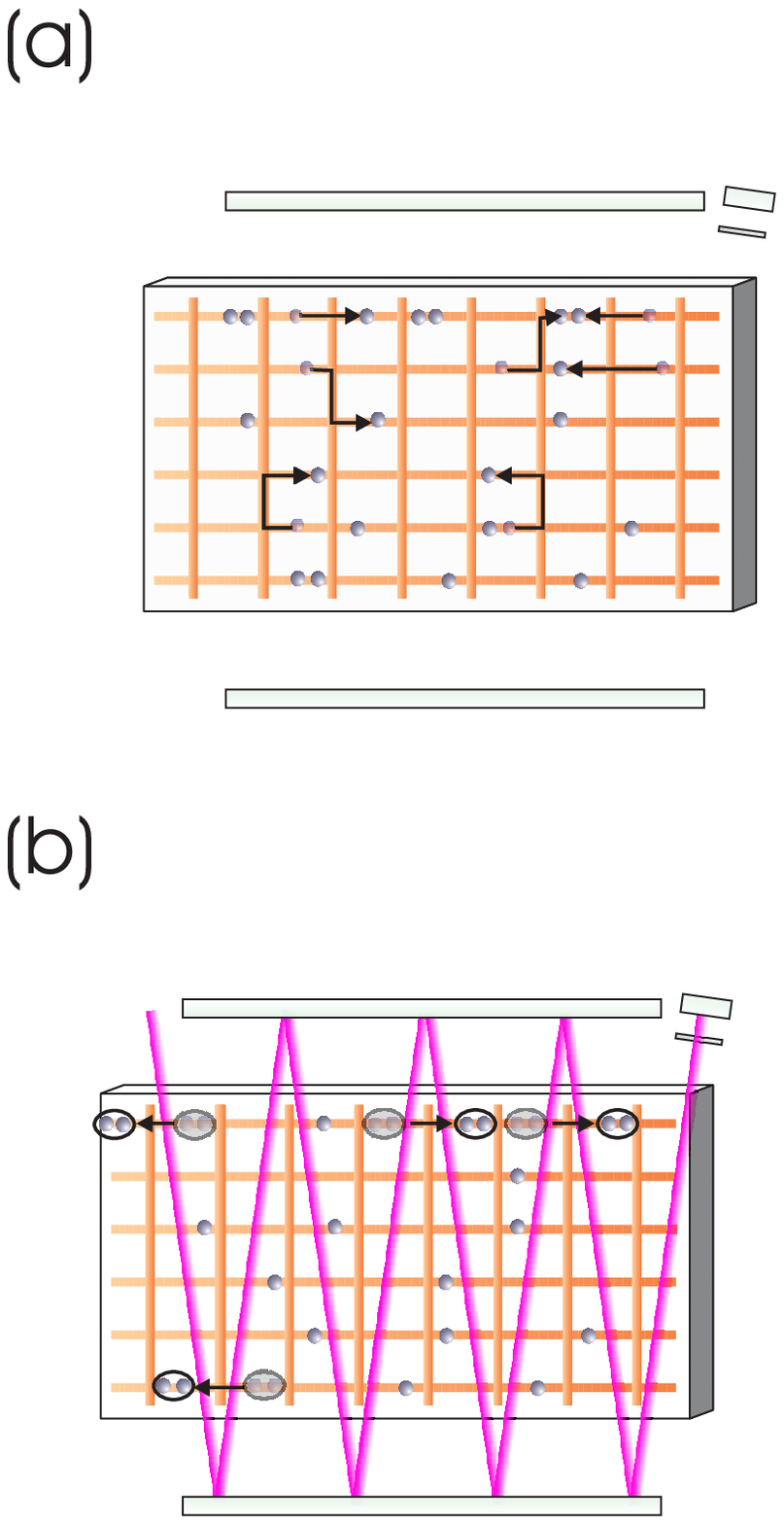}}
\caption{\label{Fig:BasArc} (Color online) Basic steps for the
proposed architecture: (a) The ions carrying the quantum information
are arranged into the desired spatial configuration in the trap
array while the laser beams are switched off. (b) All laser-beam
assisted operations scheduled after the prearrangement in (a) are
carried out. This includes one-qubit rotations, two-qubit gates and
measurements.
 }
\end{figure}

Our proposed architecture is based on a multi-zone geometry
\cite{wineland98,kielpinski02,steane04,kim05,reichle06,hucul07}. To
be specific we consider planar surface electrode trap arrays
\cite{chiaverini05} in the following, but the basic ideas should
also work in other types of trap arrays. In this architecture, logic
operations are implemented by two basic primitive steps:

\textbf{(i)} The ions carrying the quantum information are arranged
into a particular spatial configuration in the trap array while the
laser beams are switched off [Fig. \ref{Fig:BasArc} (a)].

\textbf{(ii)} All laser-beam assisted operations scheduled for the
configuration are implemented after (i) is carried out. The laser
beam(s) are collectively switched on, then single qubit ions or
pairs of qubit ions are transported through the laser beams to
implement one-qubit rotations, two-qubit gates and measurements
[Fig. \ref{Fig:BasArc} (b)]. Finally the laser beams are switched
off collectively.

Steps (i) and (ii) are repeated until the computation is finished.
In more detail, the control for ion motion in (i) can be
accomplished with a few sequential elementary sub-steps.  For
example, these sub-steps could be translations of the potential
wells containing an ion or ions in the array and splitting and
recombing potential wells to reconfigure ions into different
groupings \cite{rowe02,barrett04}.  These basic operations are
indicated schematically in Fig. \ref{Fig:BasArc} (a). In addition to
these classical means of transport, quantum information can be
transported in the array without physically moving the information
carriers by teleportation
\cite{bennett93,gottesman99a,riebe04,barrett04}. Teleportation could
be supported by a backbone of entangled qubits distributed over the
whole array before and/or in parallel with the computation. Such an
entanglement backbone could also be part of an efficient error
correction scheme \cite{knill05a}.

After the preconfiguration of qubits in the array, step (ii) is
implemented. This step can be broken down into three basic laser
assisted sub-operations that we call {\it transport gates}:
single-qubit rotations, two-qubit gates and measurement.
Single-qubit rotations can be initiated by first turning on specific
laser beams globally over the entire array. Then the qubits
scheduled for one qubit-rotations are transported through the laser
beams in a controlled fashion (see Fig. \ref{Fig:BasArc} (b)). Next
the beams for two-qubit gates are turned on and the pairs of ions
scheduled for two-qubit gates are transported through the beams.
Finally all qubits scheduled for measurement are read out by either
turning on a detection beam at their current location or
transporting them through a globally switched detection beam.
Depending on the exact nature of the detection scheme, all
measurements can be done in parallel if position-resolving detectors
are used. Alternatively, detection could be accomplished serially
with scheduled transports if no (or only limited) position
resolution is available. The physical implementation of one-qubit
rotations and two-qubit gates is discussed in sections
\ref{Sec:OneQub} and \ref{Sec:TwoQub} respectively.

Previously, temporal control of ions' internal states, such as qubit
rotations, had to be achieved by individually calibrated, precise
switching of  laser beams. With transport gates laser beams can be
switched on and off globally over the entire trap array while
precise individual control is now transferred to the ion motion.
This also facilitates the use of one set of laser beams for parallel
operations on ions distributed over the trap array, reduces the
complexity of optics, and might lead to lower requirements on the
total laser power necessary to run processors with a certain number
of qubits.

By repetition of (i) and (ii) we can realize one-qubit rotations,
two-qubit gates and measurement between arbitrary qubits in an
arbitrary order, which is sufficient for universal quantum
computation \cite{barenco95}. Individual operations are then
controlled by the motion of ions alone, while the switching of
lasers can be implemented with reduced timing precision. This could
be of significant practical importance, if active feedback on the
lasers to counteract intensity or beam position fluctuations is
desired. In the traditional scheme where operations depend on the
temporal characteristics of the laser beams, such feedback would
have to act on timescales much shorter than that of the laser
pulses. In the scheme discussed here, the transport of the ions can
be delayed by a suitable amount of time for the feedback to settle.
Such a procedure could also alleviate the detrimental effects of
other switching imperfections such as phase chirps in acousto-optic
modulators. As a further example of the potential simplification,
consider the problem of implementing in parallel a specific rotation
on several spatially separated qubits with the same laser beam.  If
the rotation is implemented by applying a pulse to ions already in
place, we require the laser intensity to be the same at the site of
each ion, a difficult task, given the general divergence/convergence
of the laser beams.  This problem can be alleviated by controlling
the transport of each qubit through the beam separately.
Furthermore, since the ions are transported completely through the
laser beams, the gate interaction does not change if beams have
small pointing instabilities in the plane of the trap array that
change on a timescale long compared to the gate interactions
(typically fulfilled for beam-steering timescales in the
laboratory). It is therefore sufficient to stabilize the beam
pointing in the direction perpendicular to the motion of the ions.

\section{Two-photon stimulated Raman interactions}\label{Sec:RamTra}

In this section we briefly review the basic interactions that play a
role in the transport gates discussed later. We assume that we have
perfect control of the ion motion in the array; in particular, we
assume that we can confine one or two qubit ions in one potential
well with precisely defined motional frequencies, and we also assume
we can translate the well along a predefined trajectory without
changing the motional frequencies. We assume that the ions start in
and remain in the motional ground state in the accelerating phases
at the beginning and end of the transport and that the well is
translated with a constant velocity $\mathbf{v}$ relative to the
laboratory frame while the ions move through the laser beams. This
situation, as viewed from the ions' frame of reference, is
equivalent to that of ions in the ground state in a stationary
potential well under the influence of temporally controlled laser
beams. We can therefore describe one-qubit gates and two-qubit gates
with the interaction Hamiltonians that are also appropriate for ions
at rest in the laboratory frame. The difference is that, for each
operation, the coordinates of the ions are transformed by
$\mathbf{r}\rightarrow \mathbf{r}_{\rm lab}+\mathbf{v}~ t$ with
$\mathbf{r}_{\rm lab}$ a point at rest in the laboratory frame that
is taken to coincide with $\mathbf{r}$ at time $t=0$. We assume the
qubit state space is spanned by two hyperfine ground states of each
ion, formally equivalent to the two states of a spin-1/2 magnetic
moment in a magnetic field. We therefore label the qubit states
$\ket{\uparrow}$ and $\ket{\downarrow}$ and denote their energy
difference by $\hbar \omega_0$ . We consider stimulated Raman
transitions \cite{monroe95,wineland98}, implemented with two laser
beams with wavevectors $\mathbf{k}_1$ and $\mathbf{k}_2$ and
frequencies $\omega_1$ and $\omega_2$. After adiabatic elimination
of the off-resonant levels, the effective interactions of one ion
with two Raman laser beams can be described in the rotating wave
approximation by the Hamiltonian
\cite{wineland98,leibfried03a,wineland03}
\begin{eqnarray}\label{GenHam}
    H_I&=&
    \hbar \Omega_0 e^{-i([\delta_0-\omega_0] t + \phi)}
    e^{i \Delta \mathbf{k}\cdot \mathbf{r}} | \uparrow\rangle \langle \downarrow|+
    {\rm h.c.}
    \nonumber \\
    &+& \hbar \Omega_\uparrow e^{-i(\delta_0 t + \phi)}
    e^{i \Delta \mathbf{k}\cdot \mathbf{r}} | \uparrow\rangle \langle \uparrow|+
    {\rm h.c.}
    \nonumber \\
    &+& \hbar \Omega_\downarrow e^{-i(\delta_0 t + \phi)}
    e^{i \Delta \mathbf{k}\cdot \mathbf{r}} | \downarrow\rangle \langle \downarrow|+
    {\rm h.c.}
    \nonumber \\
    &+& \hbar \Delta_\downarrow | \downarrow\rangle \langle \downarrow|+
    \hbar \Delta_\uparrow | \uparrow\rangle \langle \uparrow|,
\end{eqnarray}
where h.c. is the hermitian conjugate of the previous term, $\hbar$
is Planck's constant divided by 2$\pi$, $\delta_0=\omega_1-\omega_2$
is the frequency difference, $\Delta
\mathbf{k}=\mathbf{k}_1-\mathbf{k}_2$ the wavevector difference and
$\phi=\phi_1-\phi_2$ is the phase difference of the two laser fields
at $\mathbf{r}=0$. The above expression breaks the total interaction
down into the three parts: (i) interactions that can change the spin
and possibly the motional state at the same time (sideband
transitions) are associated with the Rabi frequency $\Omega_0$, (ii)
interactions that can change only the motional state are associated
with the Rabi frequencies $\Omega_s$ with $s\epsilon\{\uparrow,
\downarrow\},$ and (iii) pure level shifts $\Delta_s$ due to the
AC-Stark effect induced independently by each of the two laser
beams. Because the energy of the ion's internal and/or external
degrees of freedom changes in transitions of type (i) and (ii),
these interactions can be only resonantly driven by one photon from
each of the two electric fields (which also need to have an energy
difference appropriate for energy conservation in the process) while
the coupling in (iii) is mediated by two photons out of the same
electric field (we do not consider the case $\delta_0=0$). The Rabi
frequencies are given by sums over dipole matrix elements:
\begin{eqnarray}\label{RabFre}
    \Omega_0&=&\frac{1}{4 \hbar^2}\sum_{l}
    \frac{
    \langle \uparrow|\mathbf{d} \cdot \mathbf{E_2}|l\rangle
    \langle l|\mathbf{d} \cdot \mathbf{E_1} |\downarrow\rangle}{\Delta_{l1}}+
    \frac{
    \langle \uparrow|\mathbf{d} \cdot \mathbf{E_1}|l\rangle
    \langle l|\mathbf{d} \cdot \mathbf{E_2} |\downarrow\rangle
    }{\Delta_{l2}},
    \nonumber \\
    \Omega_{s}&=&\frac{1}{4 \hbar^2}\sum_{l}
    \frac{
    \langle s|\mathbf{d} \cdot\mathbf{E_2}|l\rangle
    \langle l|\mathbf{d} \cdot \mathbf{E_1} |s\rangle}{\Delta_{l1}} +
    \frac{
    \langle s|\mathbf{d} \cdot\mathbf{E_1}|l\rangle
    \langle l|\mathbf{d} \cdot \mathbf{E_2} |s\rangle
    }{\Delta_{l2}},
    \nonumber \\
    \Delta_{s}&=&\frac{1}{4 \hbar^2}\sum_{l}
    \frac{
    |\langle s|\mathbf{d} \cdot\mathbf{E_1}|l\rangle
    |^2}{\Delta_{l1}} +
    \frac{
    |\langle s|\mathbf{d} \cdot\mathbf{E_2}|l\rangle
    |^2}{\Delta_{l2}},
\end{eqnarray}
where $\mathbf{d}$ is the dipole operator (assumed to be real by
convention), $\mathbf{E}_1$ and $\mathbf{E}_2$  are the two (real)
laser field amplitude vectors, and
$\Delta_{lj}=\omega_j-(E_l-E_s)/\hbar$ is the detuning of laser
field $j$  ($j \epsilon\{1,2\}$) with respect to the near-resonant
level $|l\rangle$. Typically $|\downarrow\rangle$
($|\uparrow\rangle$) is in the ground-state S manifold with energy
$E_\downarrow$ ($E_\uparrow$) while $|l\rangle$ is one of the levels
in the first excited P manifold of the ion with energy $E_l$. In the
regime of power and detunings of interest for this work
\cite{ozeri06}, the hierarchy $\Delta_l \gg \omega_0 \gg
(\Omega_0,\Omega_s,\Delta_s)$ applies. In that case, the choices of
$\delta_0$ and $\Delta k$ determine which terms dominate the
evolution described by Eq. (\ref{GenHam}). For $\delta_0 \simeq
\omega_0$ the terms connected to process (i) are near-resonant,
leading to spin-flip transitions that can be highly independent of
the motion if, in addition, the laser beams are co-propagating,
leading to $|\Delta \mathbf{k}| \simeq 0$ (see below). For $\delta_0
\simeq \omega_m$ and $|\Delta \mathbf{k}| \simeq |\mathbf{k}|$, with
$\omega_m$ one of the motional frequencies of the ion(s), process
(ii) dominates, leading to state-dependent driving of the motion.

In both situations we would like to suppress (or at least precisely
control) the AC-Stark shifts (iii) that are inevitable in the
presence of the electric fields. In practice we can often balance
the AC-Stark shifts $\Delta_s$ caused by process (iii) for
$|\uparrow\rangle$ and $|\downarrow\rangle$ without diminishing the
Rabi frequencies $\Omega_0$ and $\Omega_s$ appreciably by a
judicious choice of the intensity and/or polarizations of
$\mathbf{E}_1$ and $\mathbf{E}_2$. For example in $^9$Be$^+$ two
polarizations close to linear and orthogonal to each other can be
used to yield $\Delta_\downarrow-\Delta_\uparrow \simeq 0$ and
$\Omega_{\downarrow} \simeq -2 \Omega_{\uparrow}$ for the states
$|\uparrow\rangle =|F=1,m_F=-1\rangle$ and $|\downarrow\rangle
=|F=2,m_F=-2\rangle$ in the $S_{1/2}$ electronic ground state
\cite{wineland03}. On the other hand, we can use only one of the
Raman beams (e.g. $\mathbf{E}_1\neq 0, \mathbf{E}_2$=0 with
$\mathbf{E}_1$ circularly polarized) that induce state-dependent
AC-Stark shifts to implement rotations around the z-axis of the
Bloch sphere of the form (up to an irrelevant global phase)
\begin{eqnarray}\label{PhiRot}
Z(\phi)\ket{\uparrow} &=& e^{i \phi} \ket{\uparrow}
\\\nonumber Z(\phi)\ket{\downarrow}&=& e^{-i \phi} \ket{\downarrow}.
\end{eqnarray}

\section{Paraxial description of the laser modes}\label{Sec:ParDes}

We assume that all laser beams can be described as having a Gaussian
TEM$_{00}$ transverse beam profile. A Gaussian beam with wavevector
$k$ and angular frequency $\omega$ propagating along the $z$-axis is
the lowest order solution of the paraxial wave equation and can be
described by \cite{kogelnik66}
\begin{equation}\label{GauBea}
    E(\mathbf{r},t)=\frac{E_0}{2} \frac{w_0}{w(z)} \exp \left[\frac{-r^2}{w(z)^2}\right]\exp\left[-i \left(k z- \omega
    t+\phi-\arctan(z/z_r)+\frac{k r^2}{2 R(z)}\right)\right]+ {\rm c.c.}
\end{equation}
with $w_0$ the beam waist and
\begin{eqnarray}\label{RalRan}
    w(z)&=&w_0\sqrt{1+(z/z_r)^2}\\ \nonumber
    R(z)&=&z(1+(z_r/z)^2)\\ \nonumber
    z_r&=&\frac{k w_0^2}{2}
\end{eqnarray}
For simplicity we consider only the situation where $|z|\ll |z_r|$,
although it is possible to generalize the formalism to the curved
wavefronts in the regions with $|z|\geq |z_r|$. In our region of
interest we can write $w(z)\approx w_0$, $\arctan(z/z_r) \approx 0$,
and $(k r^2)/(2 R(z))\approx~0$. Under these approximations the beam
simplifies to
\begin{equation}\label{SimGau}
    E(\mathbf{r},t)=\frac{E_0}{2} \exp[-r^2/w_0^2]\exp \left(-i (k z- \omega
    t+\phi)\right)+{\rm c.c.}
\end{equation}
This expression describes the salient points of the spatial
dependence of the beams in the following discussions.

\section{One-qubit rotations}\label{Sec:OneQub}

\begin{figure}[ht]
\centerline{\includegraphics[width=10cm]{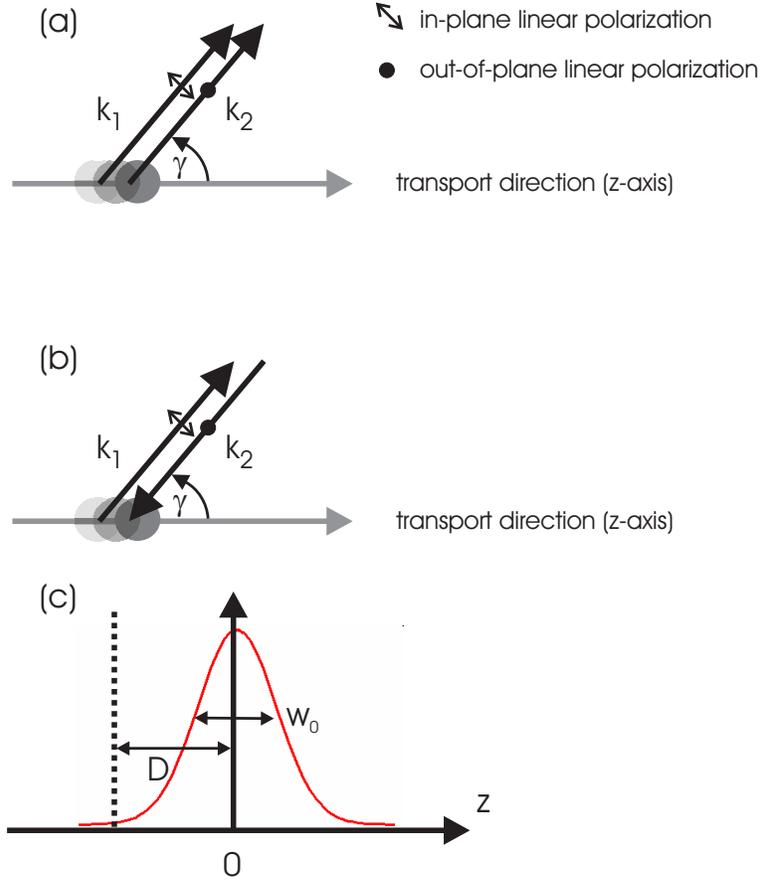}}
\caption{\label{Fig:BeaPar} (Color online) Beam parameters for one-
and two-qubit gates: (a) Orientation of the wavevectors
$\mathbf{k}_1$ and $\mathbf{k}_2$ relative to the transport
direction for co-propagating beams as used in the one-qubit
rotations. (b) Orientation of the wavevectors $\mathbf{k}_1$ and
$\mathbf{k}_2$ relative to the transport direction for
counter-propagating beams as used in the two-qubit phase gate. (c)
The starting point of the ion trajectory is a distance $D$ from the
center of the Gaussian beam profile at $z=0$ with waist size $w_0$.}
\end{figure}

Up to the present, laser-induced one-qubit rotations of trapped ion
qubits have been implemented by turning laser beams on and off for a
duration appropriate to achieve a certain rotation angle on the
Bloch sphere. By also controlling the relative phase of the laser
beams, arbitrary rotations on the Bloch sphere can be implemented.
Here we propose to move ions precisely through a laser beam to
achieve the same level of control. The duration of the interaction
is now controlled by the speed $v=|\mathbf{v}|$ of the movement. We
assume that the ion under study traverses two co-propagating laser
beams at an angle $\gamma$ relative to the $k$-vectors [see Fig.
\ref{Fig:BeaPar} (a)]. Since the wavevectors of the two laser fields
are not exactly equal, their relative phase changes by $\Delta
\phi_p \simeq s_0~ \omega_0/c=2 \pi (s_0 /\Lambda_0)$  over
locations separated by distance $s_0$ along the direction of the
beams, where $\Lambda_0$ is the wavelength corresponding to the
hyperfine transition frequency of the ion in question. Typically
$\omega_0/(2 \pi)$ is one to several gigahertz so that in
$^9$Be$^+$, $\omega_0 \simeq 2 \pi$ 1.25 GHz and $\Lambda_0 \simeq
0.24$ m. Therefore $\Delta \mathbf{k}\cdot \mathbf{r}
=(\mathbf{k}_1-\mathbf{k}_2)\cdot \mathbf{r}$ can be taken to be
constant for the typical variations of $\mathbf{r}$ of a few
nanometers, and the effective Hamiltonian Eq. (\ref{GenHam}) is
highly independent of the ion motion. If the laser frequencies
differ by $\delta_0=\omega_0$, we can drop all rapidly rotating
terms in Eq. (\ref{GenHam}) in a second rotating-wave approximation
and are left with
\begin{equation}\label{RamHam}
    H_{\rm flip}=\hbar \Omega_0 e^{-i\phi} | \uparrow\rangle \langle \downarrow|+
    {\rm h.c.}
\end{equation}
leading to Rabi rotations $R(\theta, \phi)$ on the Bloch sphere that
can be described by
\begin{eqnarray}\label{SingQub}
R(\theta,\phi)\ket{\uparrow} &=& \cos(\theta/2)\ket{\uparrow}-i e^{i
\phi} \sin(\theta/2)\ket{\downarrow}
 \\\nonumber
 R(\theta,\phi)\ket{\downarrow}&=& -i e^{-i \phi}
\sin(\theta/2)\ket{\uparrow}+\cos(\theta/2)\ket{\downarrow}.
\end{eqnarray}

For simplicity we assume that both beams have the same transverse
mode function as given by Eq. (\ref{SimGau}). The trajectory is such
that the ion is located in the center of the beams at $t=0$, where
it experiences the maximum coupling strength $\Omega_m$, which can
be calculated by inserting $E_0$ of Eq. (\ref{SimGau}) into
$\Omega_0$ of Eq. (\ref{RabFre}). The time dependent Raman-coupling
strength is then
\begin{equation}\label{CocCou}
    \Omega(t)=
    \Omega_m \exp[-2 (v \sin( \gamma) t/w_0)^2].
\end{equation}
Integration of Eq. (\ref{CocCou}) gives the rotation angle $\theta$
on the Bloch sphere resulting from a single traverse of the beams
starting at time $-T$ and ending at $T$:
\begin{equation}\label{BloAng}
    \theta = \int_{-T}^{T} \Omega(t)~ dt=
    \Omega_m \frac{w_0}{v \sin(\gamma)}\sqrt{\frac{\pi}{2}}
  ~{\rm erf}(\sqrt{2} v \sin(\gamma) T/w_0),
\end{equation}
where erf$(z) = (2/\sqrt{\pi})~\int_0^{z} e^{-t^2} dt$ is the error
function. In practice, it is permissible to neglect effects due to
the finite distance of the transport start and end points from the
center of the modes ($T\rightarrow \infty$), in which case $~{\rm
erf}(\sqrt{2} v \sin(\gamma) T /w_0)\rightarrow 1$ in Eq.
(\ref{BloAng}). The relative error $\Delta \theta/\theta$ introduced
by this approximation is $ \Delta \theta/\theta=1-{\rm erf}(\sqrt{2}
D \sin(\gamma) /w_0)$, where $D$ is the initial distance of the ion
from the center of the beams [see Fig. \ref{Fig:BeaPar} (c)]. As an
example, if $\gamma=\pi$ (beams perpendicular to the trajectory),
the infidelity $1-F \simeq (\Delta \theta/\theta)^2$ is smaller than
$10^{-4}$ if $D > 2.6~w_0$. Based on this path length of 2.6 $w_0$
we can compare the approximate duration $T_t$ to execute a transport
gate with angle $\theta$ to the duration $T_s$ necessary for the
same rotation if the ion resides in the center of beams that are
switched on and off suddenly, yielding $T_t/T_s=2.6
\sqrt{2/\pi}\simeq 2.07$.

The exact value of $\theta$ can be controlled by the speed with
which the ion is transported through the beams. For example, to
execute a $\pi$-pulse given a waist size of $w_0=$ 20 $\mu$m and
$\Omega_m=$ 250 kHz, we need a velocity $v \simeq$ 25 m/s. Similar
transport speeds (average speed 25 m/s and a peak velocity of about
45 m/s), while keeping a single ion in the ground state of the
transported potential well, have been demonstrated previously
\cite{rowe02}. Moreover, the ground state does not need to be
preserved for single qubit rotations (using co-propagating beams),
so faster, nonadiabatic transport is possible if the ion can be
sympathetically recooled after a rotation
\cite{wineland98,kielpinski02}.

The angle $\phi$ on the Bloch sphere is controlled by the relative
phase of the laser fields at $\mathbf{r}_{\rm lab}$. This phase
changes by $\Delta \phi_p \simeq 2 \pi (s_0 /\Lambda_0)$ between two
interaction zones utilizing the same beam that are separated by a
beam path of length $s_0$. Independent of whether or not temporal
laser pulses or transport gates are used, $\Delta \phi_p$ cannot be
neglected over an extended trap array. It is necessary to do careful
bookkeeping of phases over an extended trap array to properly set
and control operations. If the same pair of beams is used for
one-qubit rotations in several different places, we do not have the
freedom of setting $\phi$ independently for all these rotations.
There are several ways to deal with this problem. For example, a
straightforward, but logistically costly, solution would be to
transport a given ion to a location where $\phi$ happens to be
correct for the rotation scheduled on that ion. A more flexible
strategy would be to have a discrete set of different relative
phases that could be implemented by several beam pairs that are
turned on in parallel to address separate locations or by
sequentially switching the relative phase of the same beam
(universal computation can be achieved with a discrete set of
one-qubit rotations \cite{nielsen00}). The time overhead caused by
such sequential rotations may not slow down the computation
considerably, because one-qubit rotations are typically much faster
than two-qubit gates.

We could also momentarily turn on only one of the Raman beams.
Transporting the ion through that beam with a certain velocity
induces a time-dependent AC Stark shift that can be used to achieve
rotations $Z(\phi)$ (see the discussion preceding Eq.
(\ref{PhiRot})). A given rotation $R(\theta_2,\phi)$ with $\phi$
fixed can be turned into an arbitrary one-qubit rotation $U$ by
sandwiching it between two $Z$-operations, $U=e^{i \phi_4}
Z(\phi_3)R(\theta_2,\phi)Z(\phi_1)$ \cite{nielsen00}. In the context
of a multi-operation algorithm, this can be further simplified: The
final operation $Z(\phi_3)$ does not affect measurement outcomes and
can therefore be neglected in a one-qubit operation that precedes a
measurement. This also holds if two-qubit phase gates are applied
before the measurement since all $Z(\phi)$ commute with phase gates
\cite{footnote2}. For all other operations the final step of the
previous rotation $Z_{\rm prev}(\phi_{3})$ can be combined into the
next one-qubit manipulation $U'=R(\theta_2',\phi')Z'(\phi_1')$ where
$Z'(\phi_1')=Z(\phi_1')Z_{\rm prev}(\phi_3)=Z(\phi_1'+\phi_3)$. Thus
transporting the qubit ion through one laser beam first and then
through two beams with detuning $\omega_0$ at the appropriate
velocity is sufficient for implementing universal one-qubit
manipulations.

\section{Two-qubit gates}\label{Sec:TwoQub}

For the two-qubit gate we consider two ions trapped in the same
harmonic well. If the frequency difference of the two Raman beams is
close to a motional frequency $\omega_m$ and much smaller than the
qubit level frequency difference, $\delta_0 \ll \omega_0$, we can
neglect terms rotating with $\omega_0$ in Eq. (\ref{GenHam}). Then
the effective Hamiltonian of the ions can be approximated by
\begin{eqnarray}\label{TwoGat}
H_{zz}&=&
    \hbar \left[ \Omega_{\uparrow}(\mathbf{r}_1)| \uparrow 1 \rangle \langle \uparrow 1|+
    \Omega_{\downarrow}(\mathbf{r}_1)| \downarrow 1\rangle \langle \downarrow 1|\right]
    e^{-i(\delta_0 t + \phi)}e^{i \Delta \mathbf{k}\cdot \mathbf{r_1}} + {\rm h.c.}
    \nonumber \\
    &+& \hbar \left[\Omega_{\uparrow}(\mathbf{r}_2) | \uparrow 2\rangle \langle \uparrow 2|+
    \Omega_{\downarrow}(\mathbf{r}_2)| \downarrow 2\rangle \langle \downarrow
    2| \right]
    e^{-i(\delta_0 t + \phi)} e^{i \Delta \mathbf{k}\cdot \mathbf{r}_2}
    + {\rm h.c.},
\end{eqnarray}
where $\mathbf{r}_1$ and $\mathbf{r}_2$ are the positions of the two
ions. We use a shorthand notation for the two-qubit operators $|
\uparrow 1 \rangle \langle \uparrow 1|\bigotimes | I 2 \rangle
\langle I 2| \equiv  |\uparrow 1 \rangle \langle \uparrow 1|$ and so
on, suppressing identity operators $I$. Two ions in the same
harmonic potential well perform normal-mode oscillations around
their equilibrium position. We choose coordinates such that the axis
along which the ion conformation aligns coincides with the transport
direction ($z$-axis). The equilibrium distance is determined by the
balance of Coulomb repulsion and the restoring forces of the
external potential and is given by $d=[q^2/(2 \pi \epsilon_0 m~
\omega_{\rm COM}^2)]^{1/3}$, with $\epsilon_0$ the vacuum
permittivity, $m$ the mass and $\omega_{\rm COM}$ the center-of mass
oscillation frequency of the ions \cite{wineland98}. We can rewrite
the ion positions as $\mathbf{r}_{1}=\mathbf{r}_0 + d/2
~\mathbf{e}_z+\delta \mathbf{r}_{1}$ and
$\mathbf{r}_{2}=\mathbf{r}_0 - d/2 ~\mathbf{e}_z+\delta
\mathbf{r}_{2}$, where $\mathbf{r}_0$ coincides with the minimum of
the harmonic potential, $\mathbf{e}_z$ is a unit vector along the
$z$-axis and $\delta \mathbf{r}_{j}$ describes the small
displacements of ion $j$ around it's equilibrium position.
Substituting $\mathbf{r}_0 \rightarrow \mathbf{r}_{\rm lab} +
\mathbf{v}~t$, we can neglect the overall phase-factor $\phi+\Delta
\mathbf{k}\cdot \mathbf{r}_{\rm lab}$ that is common to both ions
and, as long as it is constant over the time of a gate interaction,
does not change the logical phase of the gate. Therefore, in
contrast to the one-qubit rotations, no bookkeeping of $\phi$ over
the trap array is necessary. We include the Doppler-shift $\Delta
\mathbf{k}\cdot \mathbf{v}t$ in $\delta_0$, define $\varphi \equiv ~
\Delta k_z d/2$ and make the Lamb-Dicke approximation, so that
$\exp[i \Delta \mathbf{k}\cdot \delta \mathbf{r}_{j}] \simeq 1+i
\Delta \mathbf{k}\cdot \delta \mathbf{r}_{j}$, which is valid
whenever the excursions $|\delta \mathbf{r}_{j}|$ are small enough
that $|\Delta \mathbf{k}\cdot \delta \mathbf{r}_{j}|\ll 1$. We also
neglect the small differences in Rabi-frequency due to the different
positions of the two ions within the Gaussian beam (assuming $d \ll
w_0$) and set $\Omega_s(\mathbf{r}_{j})\simeq
\Omega_s(\mathbf{r}_0)$ ($s ~\epsilon ~\{\uparrow, \downarrow \}$).

Using these definitions and approximations we can rewrite
Eq.(\ref{TwoGat}):
\begin{eqnarray}\label{TwoLD}
H_{zz}&\simeq&
    \hbar \left[ \Omega_{\uparrow}(\mathbf{r}_0)| \uparrow 1\rangle \langle \uparrow 1|+
    \Omega_{\downarrow}(\mathbf{r}_0)| \downarrow 1\rangle \langle \downarrow 1|\right]
    e^{-i(\delta_0 t + \varphi)}(1+i \Delta
\mathbf{k}\cdot \delta \mathbf{r}_{1}) + {\rm c.c.}
    \nonumber \\
    &+& \hbar \left[\Omega_{\uparrow}(\mathbf{r}_0) | \uparrow 2\rangle \langle \uparrow 2|+
    \Omega_{\downarrow}(\mathbf{r}_0)| \downarrow 2\rangle \langle \downarrow
    2| \right]
    e^{-i(\delta_0 t - \varphi)} (1+i \Delta
\mathbf{k}\cdot \delta \mathbf{r}_{2})
    + {\rm c.c.}
     \nonumber \\
\end{eqnarray}
We can now replace the small displacements $\delta \mathbf{r}_{j}$
by the corresponding sum of all normal-mode displacements, including
those in the directions orthogonal to $\mathbf{e}_z$. For two ions
of equal mass the two normal modes along the $z$-axis are an
in-phase (center-of-mass, COM) oscillation of both ions at frequency
$\omega_{\rm COM}$ and and out-of-phase (stretch, str) oscillation
at frequency $\omega_{\rm str}=\sqrt{3} \omega_{\rm COM}$ with
normal coordinates $z_{\rm COM} = 1/\sqrt{2}~(\delta \mathbf{r}_{1}
+ \delta \mathbf{r}_{2})\cdot \mathbf{e}_z$ and $z_{\rm str} =
1/\sqrt{2}~(\delta\mathbf{r}_{1} -
\delta\mathbf{r}_{2})\cdot\mathbf{e}_z$. To be specific, we
concentrate on detunings $\delta_0\simeq \omega_{\rm str}$ close to
the stretch mode frequency, so that $|\delta_0-\omega_{\rm str}|\ll
\{\omega_{\rm COM}, \omega_{\rm str}, \omega_{\rm str}- \omega_{\rm
COM}\}$. In the interaction picture, each normal-mode position
operator has the same form as the stretch mode operator,
\begin{equation}\label{StrOpe}
    z_{\rm str}=\sqrt{\frac{\hbar}{2 m \omega_{\rm str}}}
    (a e^{-i \omega_{\rm str} t}+ a^\dag e^{i \omega_{\rm str} t}),
\end{equation}
with $\omega_{\rm str}$ replaced by the corresponding mode
frequency. Here $a^\dag$($a$) is the harmonic oscillator mode
creation (destruction) operator. Once inserted into Eq.
(\ref{TwoLD}), all rapidly oscillating terms average to zero and are
neglected in the following; we keep the near-resonant term, which is
proportional to $e^{\pm i (\delta_0-\omega_{\rm str})}$.
Substituting the normal coordinates, setting $\eta=\Delta
\mathbf{k}\cdot \mathbf{e}_z \sqrt{\hbar/(2 m \omega_{\rm str})}$
and making the definition $\delta \equiv \omega_{\rm str}-\delta_0$
we get
\begin{equation}\label{TwLDn}
H_{zz}\simeq i \hbar \eta~\sum_{s,s'} (A_{s s'}e^{i \delta
t}~a^{\dag} -A_{s s'}^* e^{-i \delta t}~ a )|s s'\rangle\langle s
s'|,
    \end{equation}
with $s,s' \epsilon \{\uparrow, \downarrow\}$, $|s s'\rangle \langle
s s'|\equiv |s 1\rangle\langle s 1 |\bigotimes | s' 2\rangle\langle
s' 2|$ and
\begin{eqnarray}\label{StaDrv}
A_{\uparrow \uparrow} &=& - \sqrt{2} i \sin(\varphi)~
\Omega_\uparrow
\nonumber \\
A_{\uparrow \downarrow}&=& \frac{1}{\sqrt{2}}(\Omega_\uparrow
e^{-i\varphi}-\Omega_\downarrow e^{i \varphi})
\nonumber \\
A_{\downarrow \uparrow}&=&\frac{1}{\sqrt{2}}(\Omega_\downarrow e^{-i
\varphi}-\Omega_\uparrow e^{i \varphi})
\nonumber \\
 A_{\downarrow \downarrow}
&=& - \sqrt{2} i \sin(\varphi)~ \Omega_\downarrow.
\end{eqnarray}
The Hamiltonian $H_{zz}$ generates a time- and internal-state
dependent coherent drive \cite{glauber63} resulting from
state-dependent dipole forces on the ions. Since the infinitesimal
generator of the evolution $H_{zz}$ does not commute with itself for
different times, and $\delta \neq 0$, the evolution needs to be
calculated with either a time-ordered integral approach
\cite{puri01} or another method that properly evaluates the acquired
state dependent phase. As briefly described in the methods section
of Ref. \cite{leibfried03b}, we can start from the infinitesimal
displacement $d\alpha(t)= \eta A_{s s'} e^{i \delta t} $ generated
by $H_{zz}$ and calculate the total coherent displacement
\begin{equation}\label{AlpInt}
    \alpha(t)= \int_{t_0}^{t} d\alpha(t') dt'= \eta A_{s s'} \int_{t_0}^{t} e^{i \delta t'} dt',
\end{equation}
and the acquired phase
\begin{eqnarray}\label{PhaInt}
    \Phi(t) &=& {\rm Im}\left[\int_{t_0}^{t} \alpha^*(t') d\alpha(t')
    dt'\right]\nonumber \\
    &=& {\rm Im}\left[\int_{t_0}^{t} \left(\int_{t_0}^{t'} d\alpha^*(t'')~ dt'' \right)
    ~d\alpha(t') ~ dt'\right],
\end{eqnarray}
to express the evolution $U_{zz}(t) = \exp(i \Phi) \exp(i[\alpha(t)
a^\dag-\alpha^*(t) a])$ caused by $H_{zz}$. We can now restrict our
attention to the special case of two Gaussian beams of equal
frequency that have approximately orthogonal polarization and
counter-propagate with ${\mathbf{k}_1=-\mathbf{k}_2}=\mathbf{k}$ (a
polarization gradient standing wave, see Fig. \ref{Fig:BeaPar} (b)),
with angles $\gamma$ and $\pi+\gamma$, respectively, to the
transport direction of the ion pair. In the frame of reference of
the moving ions, the beams are Doppler shifted by $ \Delta \omega =
\pm |k| |v| \cos(\gamma)$, respectively, setting the relative
detuning of the beams to
\begin{equation}\label{DopShi}
    \delta_0 = 2 |k| |v| \cos(\gamma).
\end{equation}
As in the discussion before Eq. (\ref{CocCou}) we assume that both
beams have the same transverse mode function given by Eq.
(\ref{SimGau}), and the trajectory is such that the ions traverse
the center of the beams at $t=0$, where they experience the maximum
coupling strength $\Omega_{s,m}$ with $s~\epsilon~
\{\uparrow,\downarrow\}$. The time dependent Raman-coupling strength
is then
\begin{equation}\label{ConCou}
    \Omega_s=
    \Omega_{s,m} \exp[-2 (v \sin( \gamma) t/w_0)^2].
\end{equation}
We can define the approximate transit duration of the ions through
the beam $\tau \equiv w_0/(\sqrt{2} v \sin(\gamma))$. Since the
Gaussian envelope is common to both $\Omega_s$, we can rewrite Eq.
(\ref{StaDrv}) as
\begin{equation}\label{GauDrv}
A_{s,s'}(t) = A^{0}_{s,s'} e^{-t^2/\tau^2},
\end{equation}
with $A^{0}_{s,s'}\equiv A_{s,s'}(t=0)$. It is then straightforward
to solve the integral for the mode displacement in phase space Eq.
(\ref{AlpInt})
\begin{equation}\label{AlpGau}
    \alpha_{s s'}(t)= \eta A^{0}_{s s'} \int_{-t}^{t} e^{-t'^2/\tau^2+i \delta t'} dt' =
    e^{-\frac{\delta ^2 \tau^2 }{4}} \frac{\sqrt{\pi } \eta A^{0}_{s s'} \tau}{2}
     \left[\text{erf}\left(t/\tau -\frac{i
   \delta \tau}{2}\right)+\text{erf}\left(t/\tau +\frac{i
   \delta \tau}{2}\right)\right].
\end{equation}
For the final displacement, after the ions have traversed a distance
equal to several beam-waists we can evaluate Eq. (\ref{GauDrv}) at
$t=\infty$ and find
\begin{equation}\label{AlpInf}
    \alpha_{s s'}(\infty)= e^{-\frac{\delta ^2 \tau^2}{4}} \sqrt{\pi } \eta  A^{0}_{s s'} \tau .
\end{equation}
This expression shows that the end point of the trajectory can be
$\alpha_{s s'}(\infty) \neq 0$. However, by choosing $\delta
\tau/\sqrt{2}$ large we can bring the end point exponentially close
to zero. This is important since we need the motional state of the
ions to be disentangled from the internal states after the gate is
executed. The simplest way to achieve this is to make sure that the
final displacement for all combinations of internal states is zero,
or very close to zero. We return to this point and its implications
for the gate fidelity below.

It is also worth noting that from Eq. (\ref{AlpGau}), $\alpha_{s
s'}(\infty)$ is proportional to the Fourier transform of the pulse
envelope at frequency $\delta$ . This implies that we can make the
final displacements very small for any choice of envelope that has
small Fourier components at $\delta$, and the transport gates also
work for beam shapes that deviate from Gaussian as long as they are
reasonably smooth.

\begin{figure}[ht]
\centerline{{\includegraphics[width=16cm]{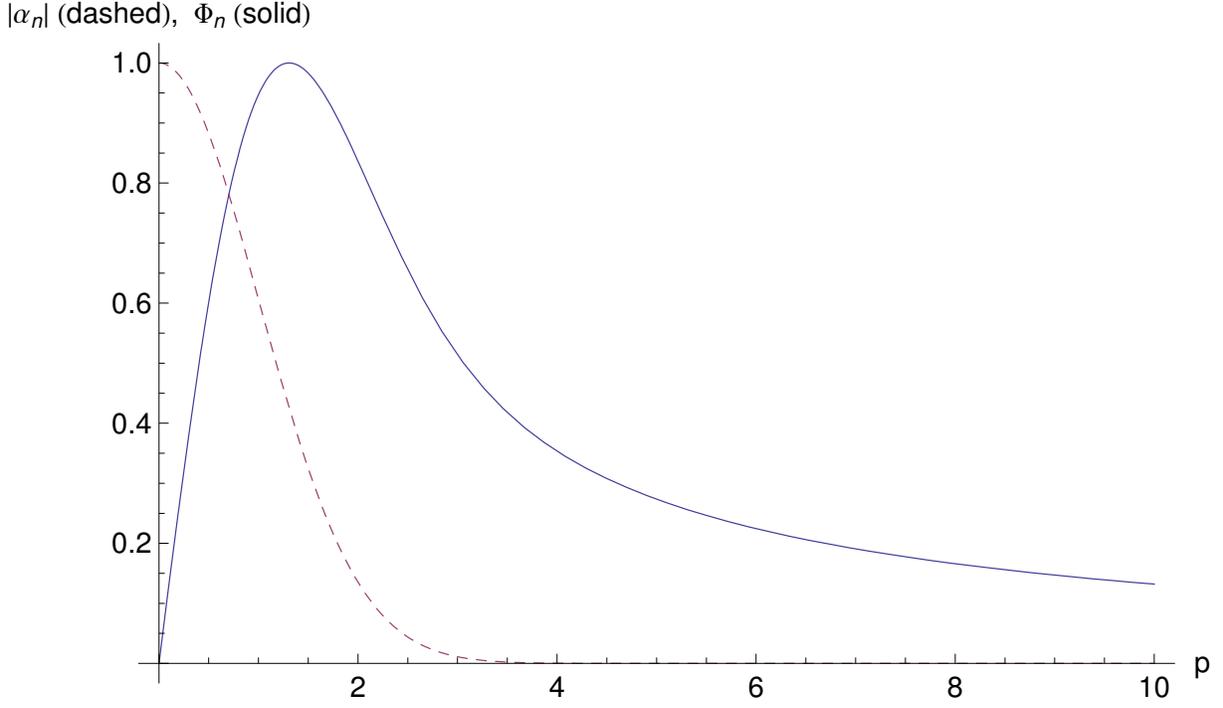}}}
\caption{\label{Fig:AlpPhi} (Color online) Normalized modulus of the
final coherent state amplitude $\alpha_n=|\alpha_{s
s'}(\infty)|/|\alpha_{s s'}(0)|$ (dashed) and normalized phase
$\Phi_n=\Phi_{s s'}/\max[\Phi_{s s'}]$ (solid) as a function of
$p$.}
\end{figure}

For the logic phases after a complete transit through the beam we
need to solve the integral
\begin{eqnarray}\label{PhaGau}
    \Phi_{s s'} &=&|\eta A^{0}_{s s'}|^2{\rm Im}\left[ \int_{-\infty}^{\infty}
    \left(\int_{-\infty}^{t'} e^{- t''^2/\tau^2-i \delta t''}~ dt'' \right)
    ~e^{-t'^2/\tau^2+i \delta t'} ~ dt'\right] \nonumber \\
    &=& |\eta A^{0}_{s s'}|^2{\rm Im}\left[ \int_{-\infty}^{\infty}
    \int_{-\infty}^{t'} e^{- t''^2/\tau^2-i \delta t''-t'^2/\tau^2+i \delta t'} ~ dt''~ dt'\right].
\end{eqnarray}
A coordinate transformation $t''=(u-v)/\sqrt{2}$,
$t'=(u+v)/\sqrt{2}$ lifts the interdependence of the integrals,
leading to a closed form solution
\begin{eqnarray}\label{TraGau}
    \Phi_{s s'} &=&|\eta A^{0}_{s s'}|^2{\rm Im}\left[
    \int_{-\infty}^{\infty}e^{-u^2/\tau^2}~du
    \int_{0}^{\infty} e^{-v^2/\tau^2+i \sqrt{2} \delta v}~ dv \right]
     \nonumber \\
    &=& |\eta A^{0}_{s s'}|^2~\tau^2 \frac{ e^{-\frac{\delta ^2 \tau^2}{2}} \pi~ \text{erfi}\left(\frac{\delta\tau}{\sqrt{2}
   }\right)}{2},
\end{eqnarray}
where $\text{erfi}(z)=-i~ \text{erf}(i z)$. Figure \ref{Fig:AlpPhi}
shows $\alpha_{s s'}(\infty)$ and $\Phi_{s s'}$, both normalized to
their respective maximal values as a function of the parameter $p
\equiv \delta \tau/\sqrt{2}$. It gives insight into the trade-off
necessary to meet the requirement $|\alpha_{s s'}(\infty)|\simeq 0$.
While $\alpha_{s s'}(\infty)$ decays exponentially in $p^2$,
$\Phi_{s,s'}$ decays more slowly with $p$ due to its dependence on
the complex error function, so that even for very small values of
$|\alpha_{s s'}(\infty)|$ the value of $\Phi_{s s'}$ remains at an
appreciable fraction of its maximal value. Note that we could also
use two well aligned transits through two beam pairs (or the same
beam pair). The acquired gate phase would add for the two passes,
while the displacements can be arranged to be opposite, so the
motional state returns to the origin. However, for this approach to
be viable the timing would have to be controlled to a small fraction
of $2 \pi/\delta$, which seems imposing, while a sufficiently small
$|\alpha_{s s'}(\infty)|$ turns out to be easily achieved for
reasonably large $p$ (see below). The resulting phase gate differs
by only one-qubit $z$-rotations from the standard phase gate
\begin{equation}\label{GatFor}
    U(\Phi_L)= | \uparrow \uparrow \rangle \langle \uparrow \uparrow|+
    | \uparrow \downarrow \rangle \langle \uparrow \downarrow|+
    | \downarrow \uparrow \rangle \langle \downarrow \uparrow|+
    e^{i \Phi_L} | \downarrow \downarrow \rangle \langle \downarrow \downarrow|,
\end{equation}
and the total logical phase of the gate can be written as
$\Phi_L=\Phi_{\uparrow \uparrow}+\Phi_{\downarrow \downarrow}
    -(\Phi_{\uparrow \downarrow}+\Phi_{\downarrow \uparrow})$
\cite{sasura03}. We find
\begin{equation}\label{TotPha}
    \Phi_L=
     -\pi e^{-\frac{\delta ^2 \tau^2}{2}}~\text{erfi}\left(\frac{\delta \tau}{\sqrt{2}
   }\right)~\frac{1}{2} \eta^2(\Omega_\uparrow - \Omega_\downarrow)^2 \tau^2 \cos (2 \varphi
   ).
\end{equation}
This expression shows that the largest modulus of the logical phase
is achieved for $\cos (2 \varphi)=\pm 1$, corresponding to $ \Delta
k_z d = n \pi$ with $n$ integer. Therefore the ion spacing should be
a multiple of the half-period of the polarization gradient wave
along the transport direction $z$. From now on we assume the ion
distance is adjusted (by a proper choice of the curvature of the
transported harmonic well) to yield a maximum modulus for the
logical phase \cite{leibfried03b,leibfried04}. This implies that
$A^{0}_{\uparrow \uparrow}=A^{0}_{\downarrow \downarrow}=0$ and
$|A^{0}_{\uparrow \downarrow}|^2=|A^{0}_{\downarrow
\uparrow}|^2=1/2(\Omega_\uparrow-\Omega_\downarrow)^2$. For a gate
equivalent to a $\pi$-phase gate we then require $\Phi_L=-\pi$,
equivalent to
\begin{equation}\label{PhaCon}
 e^{-\frac{\delta ^2 \tau^2}{2}}~ \frac{1}{2}\eta ^2 (\Omega_
  \uparrow-\Omega_\downarrow)^2 \tau^2 ~\text{erfi}\left(\frac{\delta
  \tau}{\sqrt{2}}\right)=1.
\end{equation}
With this condition, and assuming that the trajectory of the ions
starts at $\alpha_{\uparrow \downarrow}(t=-\infty)=0$, we can derive
an expression for the phase space trajectory $\alpha_{\uparrow
\downarrow}(t)$ that depends only on $p$ \cite{footnote}:
\begin{equation}\label{AlpTim}
  \alpha_{\uparrow \downarrow}(t)= \frac{\sqrt{\pi } \left(\text{erf}\left(t/\tau -\frac{i p}{\sqrt{2}}\right)
  +1\right)}{2 \sqrt{\text{erfi}(p)}}.
\end{equation}

If we assume that the only source of error in the gate is the
incomplete return to the origin for states that acquire a phase, the
fidelity of a gate can be bounded from below by the worst-case
overlap of the final state with the trap ground state,
\begin{equation}\label{GatFid}
F \geq \min_{\{c_{s s'}\}}| \sum_{s, s'} c_{s s'}~\langle \alpha_{s
s'}(\infty)|0\rangle|^2 = \min_{\{c_{s s'}\}} |\sum_{s, s'} c_{s,s'}
e^{-|\alpha_{s s'}(\infty)|^2/2}|^2 = e^{-|\alpha_{\uparrow
\downarrow }(\infty)|^2} \simeq 1-|\alpha_{\uparrow \downarrow
}(\infty)|^2,
\end{equation}
where $c_{s s'}$ are the amplitudes for state $|s s'\rangle$ in an
arbitrary two-qubit spin state the gate is operating on. Using Eq.
(\ref{PhaCon}) and Eq. (\ref{AlpGau}) we find
\begin{equation}\label{EpsCon}
   \varepsilon=1-F \leq \pi/\text{erfi}(p).
\end{equation}
We assume that subsequent gates are implemented after sympathetic
re-cooling to the motional ground state, preventing coherent
addition of errors of this type.

\begin{figure}[ht]
\centerline{\includegraphics[width=18cm]{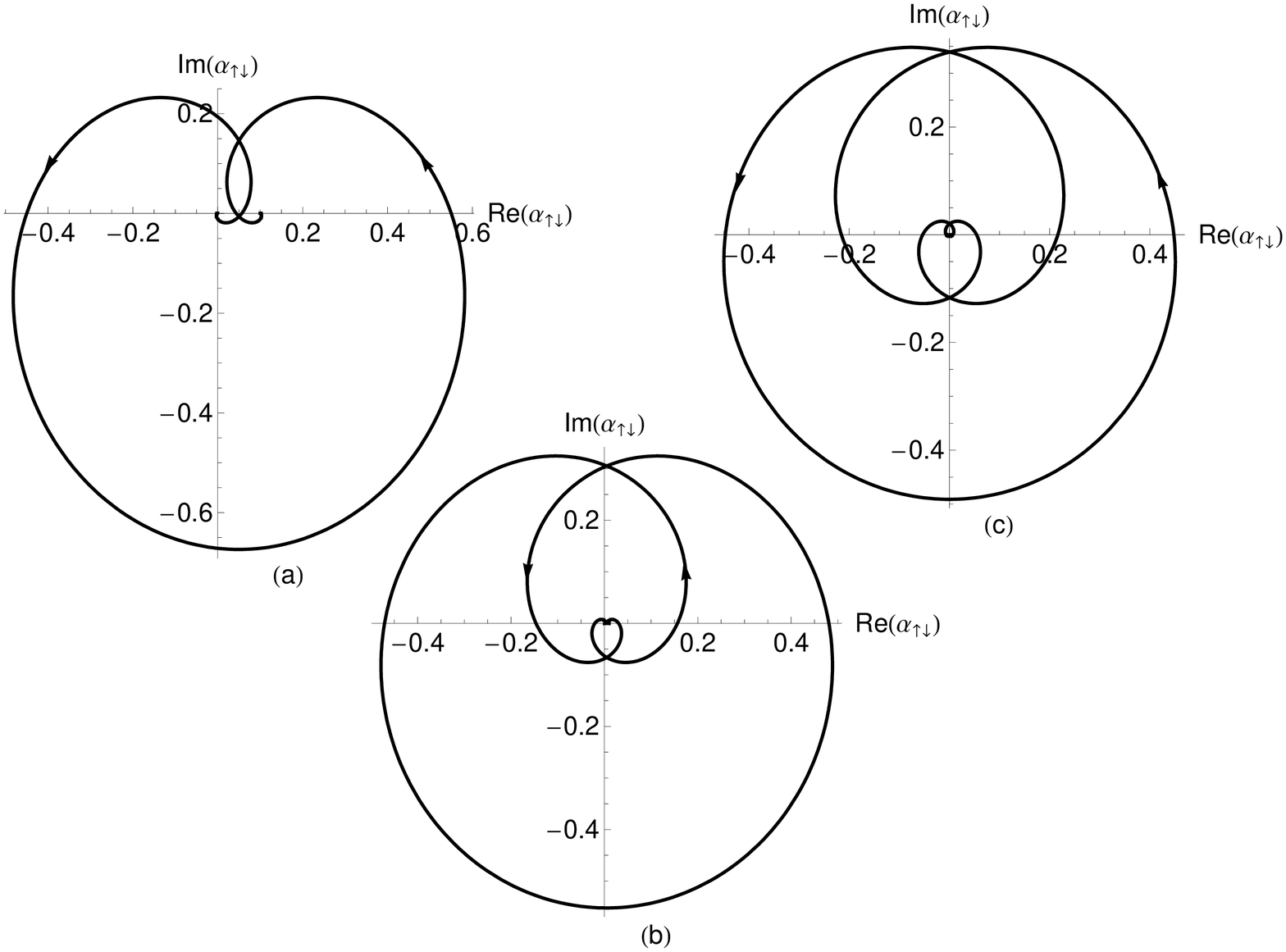}}
\caption{\label{Fig:PhaTra} Phase space trajectories of the state
$|\uparrow \downarrow \rangle$ for (a) $\varepsilon \leq 10^{-2} $
($p=2.69 $), (b) $\varepsilon \leq 10^{-4}$ ($p= 3.48$) and (c)
$\varepsilon \leq 10^{-6}$ ($p=4.11$). The arrows indicate the
trajectory directions. The trajectories are fully determined (up to
their orientation in phase space) by $p$ and the requirement that
the logical phase is $\Phi_L=-\pi$.}
\end{figure}

The trajectories in phase space of the $|\uparrow
\downarrow\rangle$-state, as described by Eq. (\ref{AlpTim}) for
phase gates with $\varepsilon \leq 10^{-2} $ ($p=2.69 $),
$\varepsilon \leq 10^{-4}$ ($p= 3.48$) and $\varepsilon \leq
10^{-6}$ ($p=4.11$), are shown in Fig. \ref{Fig:PhaTra}. For large
$|t|$, the trajectory spirals near the origin and the number of
windings per time interval $\tau$ increases as $p$ is increased. Any
desired proximity to the origin can be realized by making $p$ large
enough. We can therefore also think of $p$ as a gate-adiabaticity
parameter, because no motional energy is deposited in the system for
$p \rightarrow \infty$.

In practice it is interesting to note that $\delta_0 \tau/\sqrt{2} =
k w_0 \cot(\gamma)$ is independent of the transport speed and is
basically a measure of how many periods of the polarization gradient
wave are encountered by the ions while they are traveling through
the central portion of the beam. Unless the beam is very strongly
focussed, we have $k w_0 \gg 1$, and the ions sample many periods of
the polarization gradient wave. By choosing $\gamma$ close to 90
degrees We can reduce the number of periods encountered, but there
are drawbacks to such a choice. On the one hand, $\eta \propto
\cos(\gamma)$ becomes small. As can be seen from Eq. (\ref{PhaCon})
a smaller $\eta$ has to be compensated somehow, for example, by a
higher Rabi rate that requires more laser power, could lead to
excitation of off-resonant terms in Eq.(\ref{TwoLD}) and cause
increased decoherence from spontaneous emission \cite{ozeri06}. In
addition, to fulfill the requirement $\cos(2 \phi)=\pm 1$, we must
restrict the angle $\gamma$ to discrete values given by
\begin{equation}\label{BeaAng}
    \Delta k_z d = 2 k \cos(\gamma_n) [q^2/(2
\pi \epsilon_0 m \omega_{\rm COM}^2)]^{1/3}= n\pi,
\end{equation}
with $\gamma$ closest to $\pi$ for $n=1$. We can still choose
different values for $\gamma$ by not requiring $\cos(2 \phi)=\pm 1$,
but at the price of achieving a less than optimal logical phase.

To assess the practical feasibility of such a gate, we assume
parameters for two $^9$Be$^+$ ion qubits that are achieved in
current experiments at NIST, namely an axial trap frequency of
$\omega_{\rm COM} = 2 \pi~\times$ 4 MHz, a beam waist of $w_0=$ 20
$\mu$m, $\Omega_\uparrow=- 1/2 \Omega_\downarrow$, and a wavelength
of 313 nm. We further impose $p= 3.48$ to ensure that $\varepsilon
\leq 10^{-4}$. For $n<10$ the necessary detuning $\delta$ ranges
from 29.63 MHz to 768 kHz, too large to reasonably fulfill
$|\delta|\ll \omega_{\rm COM}$; therefore only a more complete
theory that takes the faster rotating terms into account in
Eq.(\ref{TwoLD}) would yield meaningful results. As table
\ref{Tab:Rel} shows, the transport speed $v_n$ is reasonable for all
$n \geq 10$, and gates with duration $\tau_n$ on the order of 1
$\mu$s can be realized if enough laser power is available to produce
the Rabi frequencies $\Omega_\downarrow$. Gate times are around 5
$\mu$s, comparable to the present state of the art with phase gates,
require $\gamma$ to be between 60 and 40 degrees and
Rabi-frequencies that are close to values that have been realized in
previous experiments at NIST.

\section{Effects of Sympathetic Cooling}\label{Sec:SymRec}

An integral part of QIP with ions in a large trap array is
sympathetic cooling of the qubit ions with ``refrigerator'' ions
\cite{wineland98,kielpinski02,rohde01,blinov02,barrett03} to remove
excess motional energy after transport and to reset the motional
modes into a well defined initial quantum state. Sympathetic cooling
requires the presence of neighboring ions during gate operations;
these were not considered in the preceding sections. Although the
presence of extra ions complicates the description of two qubit
gates, it does not change the essential features of the method.
There are more normal modes of motion in the extended ion
configuration and it is more complex to properly describe the
individual motional amplitudes, especially in a configuration with
ions having considerably different masses. Despite these
complications the basic gate mechanism is almost unchanged. We
consider an ion configuration with a motional eigenmode with
frequency $\omega_{v}$ and normalized eigenvector $\mathbf{v}$. We
denote the components (amplitudes) of $\mathbf{v}$ corresponding to
the ion-qubits as $v_1$ and $v_2$ (both are real numbers) and
substitute these into Eq. (\ref{TwoLD}). In complete analogy to the
calculation following that equation, we arrive at the generalized
displacement coefficients
\begin{eqnarray}\label{SymDrv}
A_{\uparrow \uparrow} &=& (v_1 e^{- i \varphi}+v_2 e^{i \varphi})~
\Omega_\uparrow
\nonumber \\
A_{\uparrow \downarrow}&=& v_1 e^{- i \varphi}\Omega_\uparrow + v_2
e^{i \varphi} \Omega_\downarrow
\nonumber \\
A_{\downarrow \uparrow}&=&v_1 e^{- i \varphi}\Omega_\downarrow + v_2
e^{i \varphi} \Omega_\uparrow
\nonumber \\
 A_{\downarrow \downarrow}
&=& (v_1 e^{- i \varphi}+v_2 e^{i \varphi})~ \Omega_\downarrow.
\end{eqnarray}
Calculating the logical phase we obtain the analog to Eq.
(\ref{TotPha}),
\begin{equation}\label{SymPha}
    \Phi_L=
     \pi e^{-\frac{\delta ^2 \tau^2}{2}}~\text{erfi}\left(\frac{\delta \tau}{\sqrt{2}
   }\right)~ 2 v_1 v_2~\eta^2(\Omega_\uparrow - \Omega_\downarrow)^2 \tau^2 \cos (2 \varphi
   ).
\end{equation}
Again the logical phase is optimal if the spacing of the qubit ions
$d_{12}$ is such that $ \Delta k_z d_{12} = n \pi$ with $n$ integer.
The suitability of a certain mode for a phase gate can be judged by
the factor $v_1 v_2~\eta^2$. Both amplitudes should be as large as
possible, while $\eta^2 \propto 1/\omega_v$ should be not too small.
At the same time, the refrigerator ions should not have components
in v that are too small, so that sympathetic cooling is efficient.
These requirements are probably best satisfied with two ion species
of comparable mass in a configuration with reflection symmetry in
the $z$-direction (for example, ($q,r,r,q$) or ($r,q,q,r$), where
$q$ denotes a qubit and $r$ a refrigerator ion).

The normal modes are then still eigenstates of the reflection
operation around $z=0$, and therefore $v_1 = \pm v_2$. For
``stretch-type'' modes with $v_1=-v_2$, the optimal coefficients
fulfill $A^{0}_{\uparrow \uparrow}=A^{0}_{\downarrow \downarrow}=0$
and $|A^{0}_{\uparrow \downarrow}|^2=|A^{0}_{\downarrow
\uparrow}|^2=v_1^2(\Omega_\uparrow-\Omega_\downarrow)^2$;
furthermore these modes should be better protected from heating
because they couple only to field fluctuations that have an
appreciable gradient over the ion configuration \cite{king98}.

One interesting aspect of sympathetic cooling in the strongly
adiabatic regime ($p \gg 1$) was already pointed out by S{\o}rensen
and M{\o}lmer \cite{sorensen00}. Due to the small deviation of the
motional wave packets from the origin and the quick succession of
revolutions in phase space, the gate becomes more robust to photon
recoil and heating of the motion (see also \cite{ozeri06}).
Therefore such gates can be executed with reasonable fidelity as
long as the heating rate (in quanta per second) is small compared to
the rate of revolution in phase space that is set by $\delta$, more
precisely, $(dn/dt)~ T_r \ll 1$ where $T_r$ is the time to complete
one revolution in phase space. Gates can even be executed while the
ions are re-cooled sympathetically, as long as the combined rates of
phonon scattering due to heating and cooling are small compared to
$\delta$. This could be advantageous when very small trap structures
are used and the ions are in close proximity to electrode surfaces,
since the heating rate has been observed to scale strongly with the
distance to the nearest surface
\cite{turchette00,deslauriers04,deslauriers06,epstein07}.
Sympathetic cooling can ensure that the motional state stays in the
Lamb-Dicke regime, even if the unchecked heating rate would drive
the ions out of this regime during a gate operation.

\section{Extensions}\label{Sec:Ext}

Hyperfine ground states with an energy splitting that is
first-order-insensitive to fluctuations in the magnetic field have
nearly identical AC-Stark shifts, leading to
$\Omega_{\uparrow}\simeq\Omega_\downarrow$
\cite{ozeri05,langer05,haljan05}. As can be seen from Eq.
(\ref{TotPha}) only a negligible logical phase can be acquired in
this situation. This problem can be solved in different ways. If
single-qubit rotations of high fidelity are available, one can
transfer out of the field-independent manifold into states with
$\Omega_{\uparrow}\neq\Omega_\downarrow$ for the duration of the
two-qubit gate operation, followed by transferring back. The ambient
magnetic field fluctuations typically do not vary appreciably over
the gate time; therefore phase errors can be sufficiently corrected
by spin-echo techniques. The extra transfer operations are a
disadvantage, but such a scheme has the benefit that qubits that are
not scheduled for gate operations are not as readily dephased by
Stark shifts caused by residual stray light from two-qubit gate
beams, which are likely to be the most powerful laser beams used in
a working QIP device.

\begin{figure}[ht]
\centerline{\includegraphics[width=16cm]{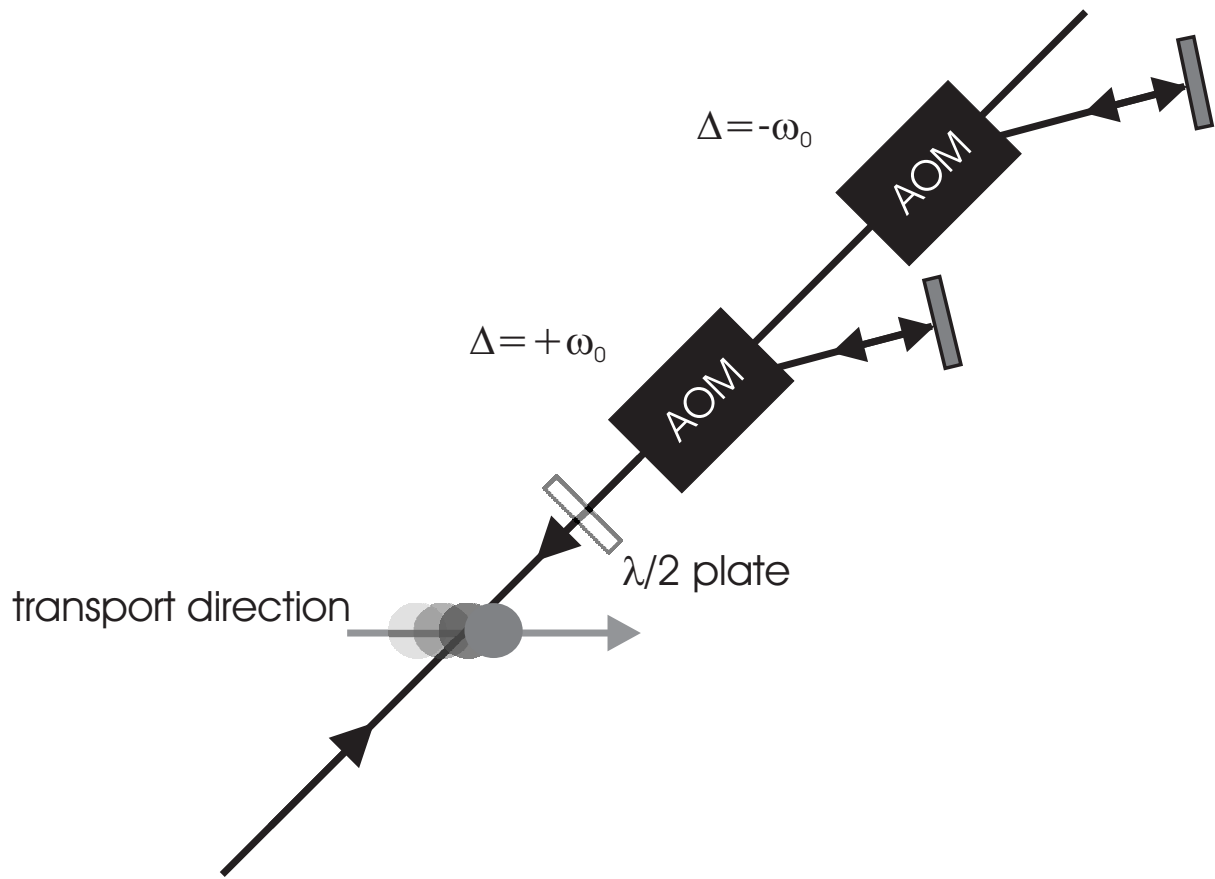}}
\caption{\label{Fig:MSmod} Possible configuraton for the transport
version of a S{\o}rensen and M{\o}lmer gate. The two double-pass
acousto-optic modulators (AOMs) shift the retro-reflected beams by
$\pm \omega_0$, respectively. Together with the Doppler shift due to
the transport, the ion experiences beat notes close to its blue and
red sidebands on the stretch mode that implement the gate.}
\end{figure}

Another approach is to use the gate first described by S{\o}rensen
and M{\o}lmer \cite{sorensen99} and Solano et al. \cite{solano99}.
The gate mechanism is basically the same as described above, but in
a rotated basis \cite{haljan05}, and can be adapted as a transport
gate in the more robust way described in Fig. 1(b) of
\cite{haljan05}. Here, two counter-propagating beams are offset from
the primary beam by $\pm \omega_0$ as shown in Fig. \ref{Fig:MSmod}.
For the proper transport velocity, this would set up a situation
equivalent to the one described in \cite{haljan05}. The frequency
differences resulting from combining these three Raman beams,
including the appropriate Doppler-shifts due to the transport, are
at $\omega_0 \pm \omega_{\rm str}+ \delta$, creating slow rotating
terms near the blue $\omega_0+\omega_{\rm str}$ and red
$\omega_0-\omega_{\rm str}$ sideband transitions in the Hamiltonian
(\ref{GenHam}). This scheme has the advantage that it works for
qubits with field-insensitive qubit transition frequencies. However,
this comes with some disadvantages. One is the added technical
difficulty inherent in creating the frequencies of the required
beams. For example, the scheme depicted in Fig. \ref{Fig:MSmod}
would require acousto-optical modulation at $\omega_0/2$. The
smallest value of $\omega_0$ for typically considered
hyperfine-state ion-qubits is $2 \pi~\times$ 1.25 GHz in $^9$Be$^+$.
Even in this case we would need two specialized modulators at
approximately 600 MHz, adding technical complexity. For
hyperfine-splittings exceeding 2 GHz the modulation would currently
require several acousto-optic and/or electro-optic modulators, or a
means to offset-phase-lock the respective laser beams. Another
disadvantage is the relative proximity of the qubit carrier
transition at the difference frequency $\omega_0$. To keep
perturbations from carrier excitation small, we have to use gate
durations much longer than  $2 \pi/\omega_{\rm str}$. For the
Z-phase gate described in the previous sections, the carrier is
essentially detuned by $\omega_0$, so gate durations comparable to
or shorter than $2 \pi/\omega_{\rm COM}$ could be possible.

Another possible extension consists of fine tuning the phase-space
trajectories with transport patterns featuring nonuniform
velocities. Such schemes could lead to better suppression of
off-resonant terms and possibly to higher gate speeds. This would
probably require numerical modeling, while the simple
uniform-velocity transport through Gaussian beams has the appeal
that all relevant quantities can be expressed analytically.

More generally the transport gate mechanism can be implemented with
any standing wave field that imparts an internal state-dependent
force. For example, close proximity of the ions to the trap
electrode surfaces should make it possible to set up strong magnetic
field gradients to induce state-dependent forces. We could construct
a magnetic ``washboard'' resulting from parallel current-carrying
wires with alternating current direction, or strips of permanent
magnetic material with alternating polarization oriented
perpendicular to the transport direction. As long as the qubit
information is contained in states with different energy dependence
(slope) as a function of the external field, the field gradients of
the washboard produces state-dependent forces. By controlling the
transport velocity of the ions we could make them experience these
forces in near resonance with a motional mode frequency. Once this
situation is established, the gate mechanism is analogous to what
has been described in the previous sections, where the strength and
envelope of the alternating force would be given by the particular
values of the currents used or the strength of the magnetized
strips. Thus a range of envelope functions could be achieved. For
example, assume a washboard made of FePt stripes (or domains written
into a continuous film) with an alternating magnetization on the
order of 800 kA/m, and a period length of $d_m \simeq$ 20 $\mu$m
could be produced (similar to those of Ref. \cite{xing06}). As a
$^9$Be$^+$ ion is transported at a height of 17.5 $\mu$m over such a
structure at $v_W =$ 80 $\mu$m/$\mu$s, this would create a
time-dependent magnetic field with an amplitude of
\begin{equation}\label{Bwa}
    B_{\rm tot}(t)=\sqrt{(B_0-B_w \cos(\omega_w t))^2+(B_w \sin(\omega_w
    t))^2} \simeq B_0(1+1/4(B_w/B_0)^2)-B_w \cos(\omega_w t),
\end{equation}
with $B_w \simeq$ 20 G, $B_0 \simeq$ 120 G \cite{hinds99}. This
choice of $B_0$ ensures first-order field independence to a qubit
encoded in the $|F=1,m_F=0\rangle$ and $|F=2,m_F=1\rangle$ states in
all regions of the array that are sufficiently far from the
permanent magnetic elements \cite{langer05}. To lowest order we have
an oscillating field at $\omega_w= 2 \pi v_w/d_m=$ 4 MHz that is
superimposed on a static average field that is larger than $B_0$ by
$B_w^2/(4 B_0) \simeq$ 0.837 G. We assume that good single-qubit
rotations are available to transfer the qubit information into the
$|F=2,m_F=-2\rangle$/$|F=2,m_F=2\rangle$ ``stretched'' states of the
hyperfine manifold. Then the undulating field of the washboard would
lead to a gate Rabi frequency of
\begin{equation}\label{MagRab}
    \Omega_m = \frac{2 \pi}{d_m} z_0 \frac{\mu_B B_w}{\hbar} \simeq
    2 \pi \times 73.5~{\rm kHz},
\end{equation}
with $\mu_B$ the Bohr magneton and $z_0 =\sqrt{\hbar/(4 m
\omega_{\rm COM})}$ the extent of the ground-state wavefunction of
the COM mode of two beryllium ions. This Rabi frequency would
correspond to a gate duration $\tau_m =\pi/\Omega_m \simeq 6.8~
\mu$s. At the same time, the slightly higher average field leads to
effective single-qubit rotations of about $2 \pi \times$ 15.9 that
commute with the phase gate. These unwanted single-qubit phases
could possibly be eliminated by a more sophisticated (symmetrical)
geometry of the elements producing the oscillating field, or by
dividing the gate into two interactions with washboards that each
yield a logical phase of $\pi/4$ and are enclosed in a spin-echo
sequence \cite{leibfried03b}.

Gate speeds are limited by the condition that the spins should
follow the magnetic field adiabatically. For this to happen, the
normalized rate of change of the field $|(d\mathbf{B}/dt)|/B_0$ has
to be much smaller than the smallest Larmor frequency along the
trajectory $\omega_{min} = \min( \gamma_< |\mathbf{B}|)$, with
$\gamma_<$ the smallest gyromagnetic ratio of the states in question
and $\mathbf{B}$ the position-dependent magnetic field. The speed of
variation of $\mathbf{B}$ is governed by the trap frequency,
therefore, at least for a gate utilizing adiabatic following, the
gate duration is limited by the trap frequency. The ``Lamb-Dicke''
parameter that is relevant for this gate is $\eta_m=(2 \pi/d_m)~
z_0$. In contrast to gates mediated by laser fields with wavelengths
prescribed by the internal states of the ions, this parameter can be
scaled more freely by changing the period length $d_m$ of the
microfabricated pattern. Of course the ions have to be moved closer
to the surface to create a sufficient modulation of the field along
their trajectory. This might lead to an increase in anomalous
heating, but at the same time the closer proximity of the ions
allows use of smaller trap electrodes and scaling the dimensions of
the trap array. For gates mediated by light, the need to illuminate
ions by laser beams with waists that are most likely limited to
sizes of several micrometers from geometrical constraints dictates a
minimum distance of the ions to the trap electrodes. This distance
in turn governs the minimal electrode dimensions in a trap array. If
light is not necessary to mediate the gates we could use trap
structures beyond this limit, possibly with stronger confinement of
the ions and faster gates.

An additional, but important advantage of such gates is the absence
of spontaneous emission that is currently thought to pose the most
severe limitation on achievable fidelities for Raman transition
gates implemented with laser beams \cite{ozeri06}.

\section{Summary and Conclusions}\label{Sec:Con}

Miniaturization of ion traps and scaling up ion qubit numbers to
beyond 10 is now under way in several laboratories. Recent efforts
have concentrated on developing technologies for trap arrays
compatible with this goal. One of the next steps is to simplify the
optics necessary for scale-up as much as possible. In this paper we
have outlined a few possible techniques towards this goal.
Controlled transport of ions is utilized for parallel implementation
of quantum logic gates with relaxed requirements for temporal and
spatial control of the laser beams. After sketching the basic
features of such an architecture based on a multi-zone trap array,
we analyzed one-qubit rotations and a universal two-qubit gate and
showed that such gates can be implemented with existing technology.

We also discussed possible extensions of the two-qubit gate
mechanism. In particular we sketched one approach that utilizes
periodic magnetic field patterns in combination with ion transport
to exert state-dependent forces on the ions. This approach might
enable more flexible scaling of future trap arrays and could remove
some of the limitations to fidelity and gate speed posed by
two-qubit gates mediated by laser beams.

\section*{Acknowledgements}
This work was supported by the Disruptive Technology Office (DTO)
under contract number 712868, by the DoD Multidisciplinary
University Research Initiative (MURI) program administered by the
Office of Naval Research and by NIST. We thank S. Glancy and T.
Rosenband for comments on the manuscript. This paper is a
contribution of the National Institute of Standards and Technology
and not subject to U.S. copyright.

\newpage
\bibliographystyle{apsrev}
\bibliography{jrnls,transportgate}

\begin{thebibliography}{45}
\bibitem[{foo({\natexlab{a}})}]{footnote2}
\bibinfo{note}{Both operations are diagonal in the measurement basis.}
\bibitem[{foo({\natexlab{b}})}]{footnote}
\bibinfo{note}{Here we neglect a sign that depends on the value of $n$ and the
  sign of $\Omega_\uparrow-\Omega_\downarrow$, but will only determine the
  orientation of the trajectory in phase space.}
\end{thebibliography}
\newpage
\begin{table}[ht]
 $
\begin{array}{lllllll}
 n ~~& \gamma /{
 \rm deg}~~ & \eta ~~~~~~~& v/{\rm(m/s)}~~ & \tau / \mu s~~ & \delta /(2 \pi  {\rm MHz})~~ &
 \Omega_\downarrow /(2 \pi  {\rm MHz}) \\
 10 & 77.6 & 0.077 & 5.50 & 1.32 & 0.596 & 3.579 \\
 12 & 75.1 & 0.093 & 4.52 & 1.62 & 0.484 & 2.423 \\
 14 & 72.6 & 0.108 & 3.83 & 1.93 & 0.405 & 1.738 \\
 16 & 70.0 & 0.124 & 3.32 & 2.26 & 0.346 & 1.300 \\
 18 & 67.3 & 0.139 & 2.94 & 2.61 & 0.300 & 1.003 \\
 20 & 64.7 & 0.155 & 2.63 & 2.98 & 0.263 & 0.791 \\
 22 & 61.9 & 0.170 & 2.38 & 3.37 & 0.233 & 0.636 \\
 24 & 59.1 & 0.186 & 2.17 & 3.79 & 0.207 & 0.518 \\
 26 & 56.2 & 0.201 & 2.00 & 4.25 & 0.184 & 0.426 \\
 28 & 53.2 & 0.217 & 1.85 & 4.77 & 0.164 & 0.353 \\
 30 & 50.1 & 0.232 & 1.72 & 5.35 & 0.147 & 0.293 \\
 32 & 46.8 & 0.247 & 1.61 & 6.02 & 0.130 & 0.245 \\
 34 & 43.3 & 0.263 & 1.51 & 6.80 & 0.115 & 0.204 \\
 36 & 39.6 & 0.278 & 1.43 & 7.77 & 0.101 & 0.168 \\
 38 & 35.6 & 0.294 & 1.35 & 9.00 & 0.087 & 0.138 \\
 40 & 31.1 & 0.309 & 1.28 & 10.69 & 0.073 & 0.110 \\
 42 & 26.0 & 0.325 & 1.22 & 13.26 & 0.059 & 0.085 \\
 44 & 19.7 & 0.340 & 1.16 & 18.13 & 0.043 & 0.059 \\
 46 & 10.1 & 0.356 & 1.10 & 36.41 & 0.022 & 0.028
\end{array}
$ \caption{\label{Tab:Rel} Gate parameters for an axial trap
frequency of $\omega_{\rm COM} = 2 \pi \times$ 4 MHz, beam waist of
$w_0=$ 20 $\mu$m, $\Omega_\uparrow=- 1/2~ \Omega_\downarrow$,
wavelength of 313 nm ($^9$Be$^+$) and $p = 3.48$ ($\varepsilon \leq
10^{-4}$). All parameters, including Rabi-frequencies
$\Omega_\downarrow$, are within experimental reach.}
\end{table}
\newpage
\newpage
\newpage
\newpage
\newpage

\end{document}